\documentclass[journal]{IEEEtran}
\IEEEoverridecommandlockouts
\usepackage{cite}
\usepackage{amsmath,amssymb,amsfonts}
\usepackage{algorithmic}
\usepackage{graphicx}
\usepackage{textcomp}
\usepackage{xcolor}
\def\BibTeX{{\rm B\kern-.05em{\sc i\kern-.025em b}\kern-.08em
    T\kern-.1667em\lower.7ex\hbox{E}\kern-.125emX}}

\usepackage{float}
\usepackage{caption}

\usepackage[font=footnotesize]{subfig}
\usepackage{graphicx}
\usepackage{dblfloatfix}    
\usepackage[hidelinks]{hyperref}

\begin{document}

\title{System Integration of Xilinx DPU and HDMI for Real-Time inference in PYNQ Environment with Image Enhancement}

\author{Jonathan Sanderson,
        Syed Hasan,~\IEEEmembership{Senior Member,~IEEE,}
}

\maketitle

\begin{abstract}
Use of edge computing in application of Computer Vision (CV) is an active field of research.
Today, most CV applications make use of Convolutional Neural Networks (CNNs) to inference on and interpret video data.
These edge devices are responsible for several CV related tasks, such as gathering, processing and enhancing, inferencing on, and displaying video data.
Due to ease of reconfiguration, computation on FPGA fabric is used to achieve such complex computation tasks. 
Xilinx provides the PYNQ environment as a user-friendly interface that facilitates in Hardware/Software system integration. 
However, to the best of authors' knowledge there is no end-to-end framework available for the PYNQ environment that allows Hardware/Software system integration and deployment of CNNs for real-time input feed from High Definition Multimedia Interface (HDMI) input to HDMI output, along with insertion of customized hardware IPs. 
In this work we propose an integration of rea\textbf{L}-time image \textbf{E}nancement IP with \textbf{A}I inferencing engine in the \textbf{P}ynq environment (\textbf{LEAP}), that integrates HDMI, AI acceleration, image enhancement in the PYNQ environment for Xilinx's Microprocessor on Chip (MPSoC) platform.
We evaluate our methodology with two well know CNN models, Resnet50 and YOLOv3.
To validate our proposed methodology, LEAP, a simple image enhancement algorithm, histogram equalization, is designed and integrated in the FPGA fabric along with Xilinx's Deep Processing Unit (DPU).
Our results show successful implementation of end-to-end integration using completely open source information. 

\end{abstract}


\section{Introduction} \label{intro}
From smart cars to security cameras, Computer Vision (CV) plays a vital role in our modern world.
This is majorly due to the use of Convolutional Neural Networks (CNNs), providing several advancements over traditional algorithms \cite{cv_survey}.
As using CNN's has become widely desired, implementing them on edged devices has been an active field of research \cite{edge_latency} \cite{adeyemo2023stain}.
CNN's require specialized hardware to achieve real-time inference.
Field Programmable Gate Arrays (FPGAs) are good hardware choice due to their reconfigurability and parallel processing logics \cite{intel-whitepaper}.
Typically, CNNs are trained on datasets of procured images.
When CNNs are used in the field, they are likely to encounter adverse conditions, that degrades their performance.
For example, a security camera in a room with minimal brightness, may not be able to properly detect the presence of people in the room.
More adverse scenarios are possible for different applications, such as hazy or rainy conditions for autonomous vehicles.
A solution to this is image enhancement \cite{image-enhance-fpga}.
An example of this is taking a dark image and brightening it, to improve CNN's capability of prediction.
Along these lines, in \cite{sandefur2022framework}, the authors provide a methodology for benchmarking CNNs on 720p video.
They, however, do not provide any methodology for implementing image enhancements.
The authors in \cite{retinex-fpga} explored using an image enhancement algorithm, Retinex \cite{retinex-algorithm}, in conjunction with You Only Look Once (YOLO) object detection model \cite{yolov2}.
While the work in \cite{retinex-fpga} provides a methodology for integrating an image enhancement and a CNN model, it lacks in the details of evaluating metrics and methodology of integrating CNN modules to their IPs. 
Authors in \cite{sanderson2023integrating} explored improving \cite{sandefur2022framework} by adding image enhancement into the Gstreamer pipeline.
However, this approach is limited due to image enhancement being implemented in software and not on the FPGA fabric.
The above discussion illustrates that currently literature is deficient in providing  a systematic approach for integrating an end-to-end system using FPGA fabric from the input video stream to output passing through image processing and CNN models. 
To achieve this integration, researchers leverage information from Xilinx's documentation database for implementing their algorithms in the FPGA fabric.
Xilinx provides documentation for their CNN accelerator IP, referred to as Deep Processing Unit (DPU) \cite{dpu-documentation} and their High-Definition Multimedia Interface (HDMI) IP \cite{hdmi-documentation}. 
Xilinx also provides an example of combining the HDMI and DPU \cite{hdmi_trd} IPs, however information on integrating of another IP into this HDMI and DPU environment  is not publicly available.
The task of inferencing a CNN on the DPU requires close-knit orchestration of hardware in software.
Complex drivers are required to communicate with dynamic hardware.
PYNQ \cite{pynq} is a runtime environment that greatly simplifies the software by providing drivers and Application Programming Interfaces (APIs), but requires integration to work correctly.
There are software examples for using HDMI \cite{pynq-hdmi-notebook} and the DPU \cite{pynq-dpu-notebook} at runtime, but to the authors' knowledge there are no guidelines available to systematically integrate HDMI in, IP, DPU, and HDMI out together in the PYNQ environment.
This leads to following research challenges. 
\\ \textbf{Research Challenges:}
\vspace{-1.5mm}
\begin{enumerate}
    \item What is the approach to integrate HDMI input, IPs, DPU, and HDMI output within an FPGA's fabric?
    \item What are the requirements of the above integration to be embedded in the PYNQ environment?
    \item How to evaluate the performance of the AI-models integrated in the above system?
\end{enumerate}
\vspace{-1mm}
We propose a methodology, integration of rea\textbf{L}-time image \textbf{E}nancement IP with \textbf{A}I inferencing engine in the \textbf{P}YNQ environment (\textbf{LEAP}), for integrating hardware based HDMI in/out, DPU, image enhancement in the PYNQ environment for the ZCU104 development platform \cite{zcu104}.
We provide details of both hardware and software integrations.
We demonstrate three different ways of inferencing the CNN on video in real-time.
To validate our methodology, we use a well known enhancement algorithm, histogram equalization \cite{image-proc-book}.
We use this enhancement algorithm to enhance dark images before they are inferenced on by two well know CNNs, Resnet50 \cite{resnet50-model} and YOLOv3 \cite{yolov3}, collecting frame rate and accuracy as metrics.
\\ \textbf{Contribution:}
\vspace{-1.5mm}
\begin{enumerate}
    \item A complete end-to-end methodology is proposed for integrating HDMI in/out, DPU, and image enhancement together in the PYNQ environment.  For promoting open research we are providing complete code to recreate our work (we'll provide our code as a weblink, and include it as a citation in this paper).
    \item An evaluation technique is developed for testing the effect of image enhancements on CNN inference.
    \item Multiple options of inferencing methodology based on their hardware organization are developed, and their results are provided.
\end{enumerate}
\vspace{-1.5mm}

\section{Platforms Used: Xilinx's Deep Learning Processing Unit (DPU) and PYNQ Environment} \label{background}

To aid the reader, we are providing a some background on topics we commonly refer to.
CNNs often have millions of parameters.
Inferencing a CNN typically requires computing billions of Multiply ACcumulate (MAC) operations.
They are often trained using graphical processing units (GPU's) with batches of images.
CNNs can be inferenced on CPUs, but for an edge computing environment this is impractical because of the compactness of edge devices.
Xilinx provides an accelerator intellectual property (IP) designed for FPGA fabric that can accelerate the inference of CNNs and can be used in reconfigurable edge devices.
They offer different sizes of their DPU IP, we are using their largest size, which is capable of computing 1228 billion operations per second.
The DPU cannot run any arbitrary CNN model, but requires a model to be translated to target the DPU architecture.
To aid research, Xilinx provides a whole suite of CNN models from their Vitis-AI model zoo \cite{model-zoo}.
We are using PYNQ as it is developer friendly for rapid prototyping environments.
PYNQ uses the Python programming language for high level control.
It provides Python libraries to control low level drivers that configure and communicate with IPs in the PL. 
It allows for rapid prototyping via the Python programming language.

\section{LEAP-Methodolgy} \label{method}

To implement LEAP, we are using Xilinx ZCU104 development board \cite{zcu104}. 
The ZCU104 is an FPGA that combines four ARM processors and programmable logic into a microprocessor on chip (MPSoC). 
This MPSoC has an ARM processor (called as PS by Xilinx) and FPGA fabric (called as PL or programming logic by Xilinx).
The PS and PL interface work closely together. 
The PL can be reconfigured at run-time with precompiled bitstreams.
These bitstreams contain IPs that then can be used to accelerate programs by running them in the PL.
The configuration of the ZCU104 is further explained in the following three subsections: the PL and bitstream generation, compiling the PYNQ image for booting, and runtime software.

\subsection{Program Logic and Bitstream Generation} \label{program-logic}

\begin{figure}[h]
    \centering
    \includegraphics[width=0.48\textwidth]{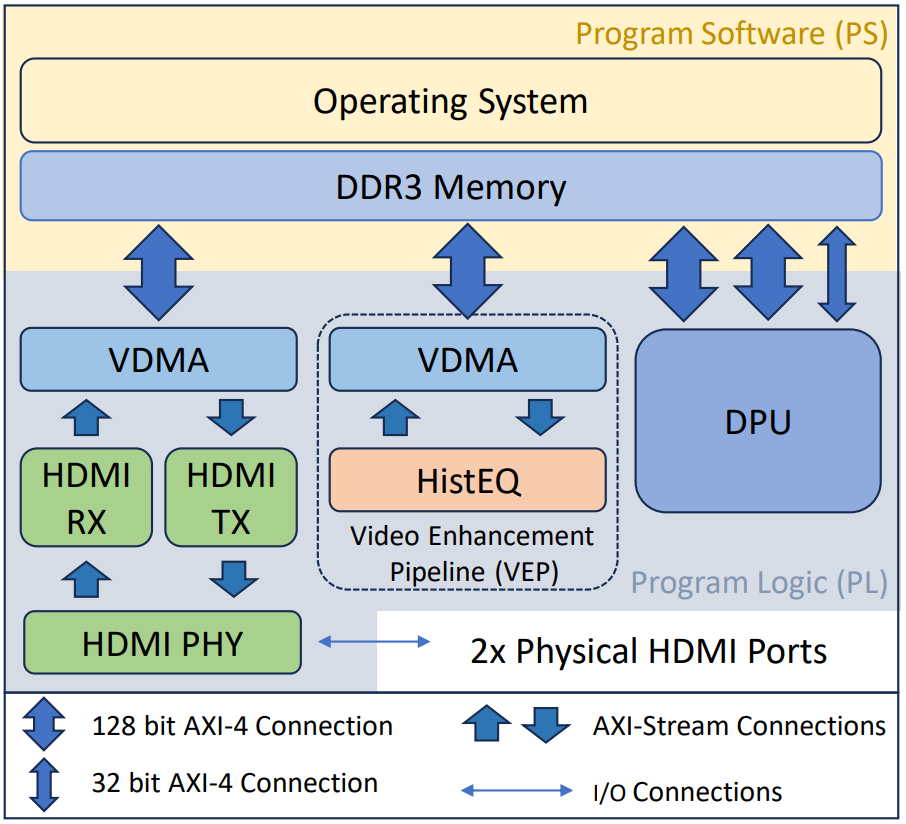}
    \caption{System overview. Part of our contribution is the integration of the Video Enhancement Pipeline (VEP).}
    \label{fig:sys-overview}
\end{figure}

\begin{figure}[H]
    \centering
    
    \subfloat[\label{fig:histeq-example:a}]{\includegraphics[width=0.49\linewidth]{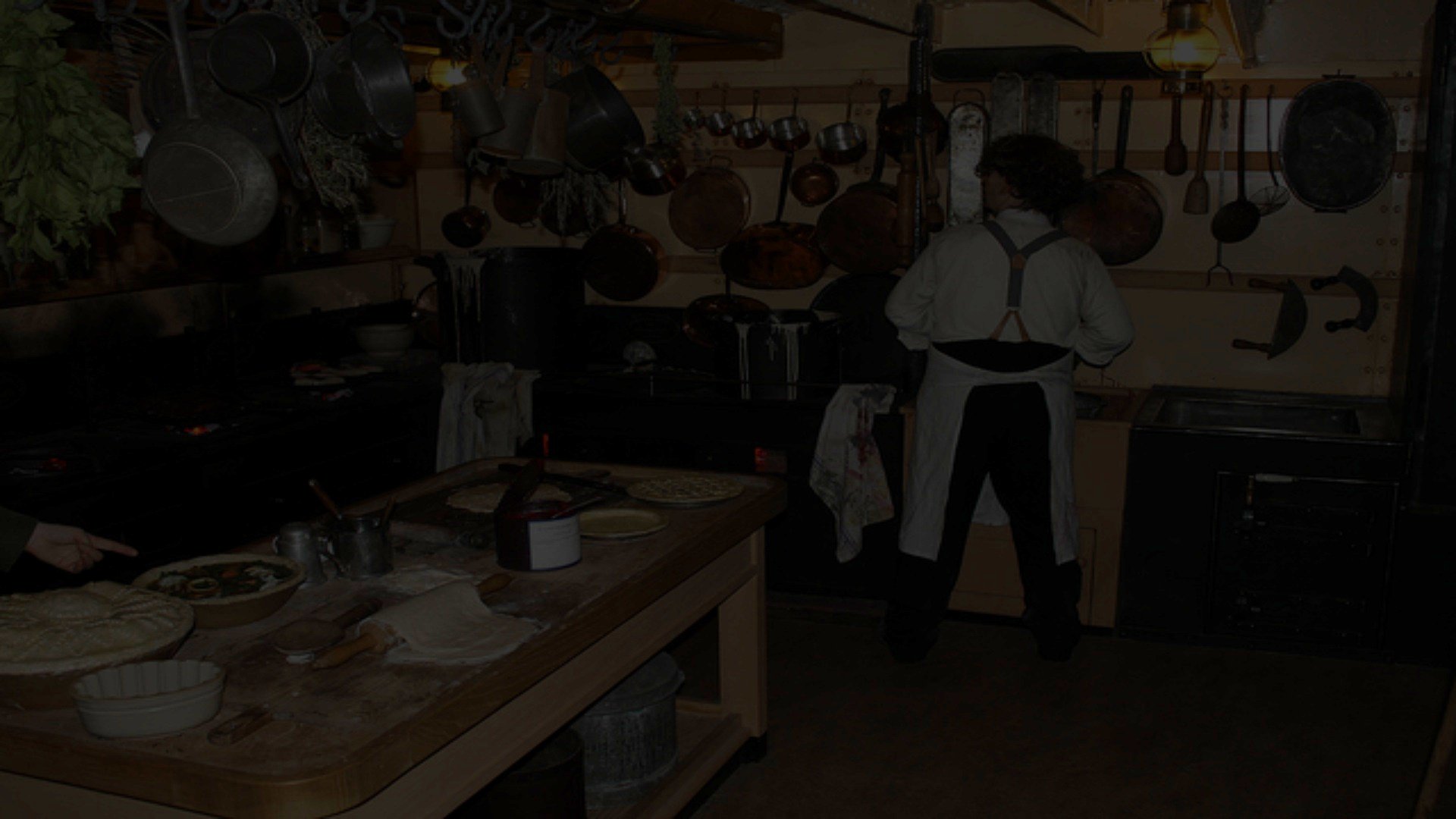}}
    \hfill
    \subfloat[\label{fig:histeq-example:b}]{\includegraphics[width=0.49\linewidth]{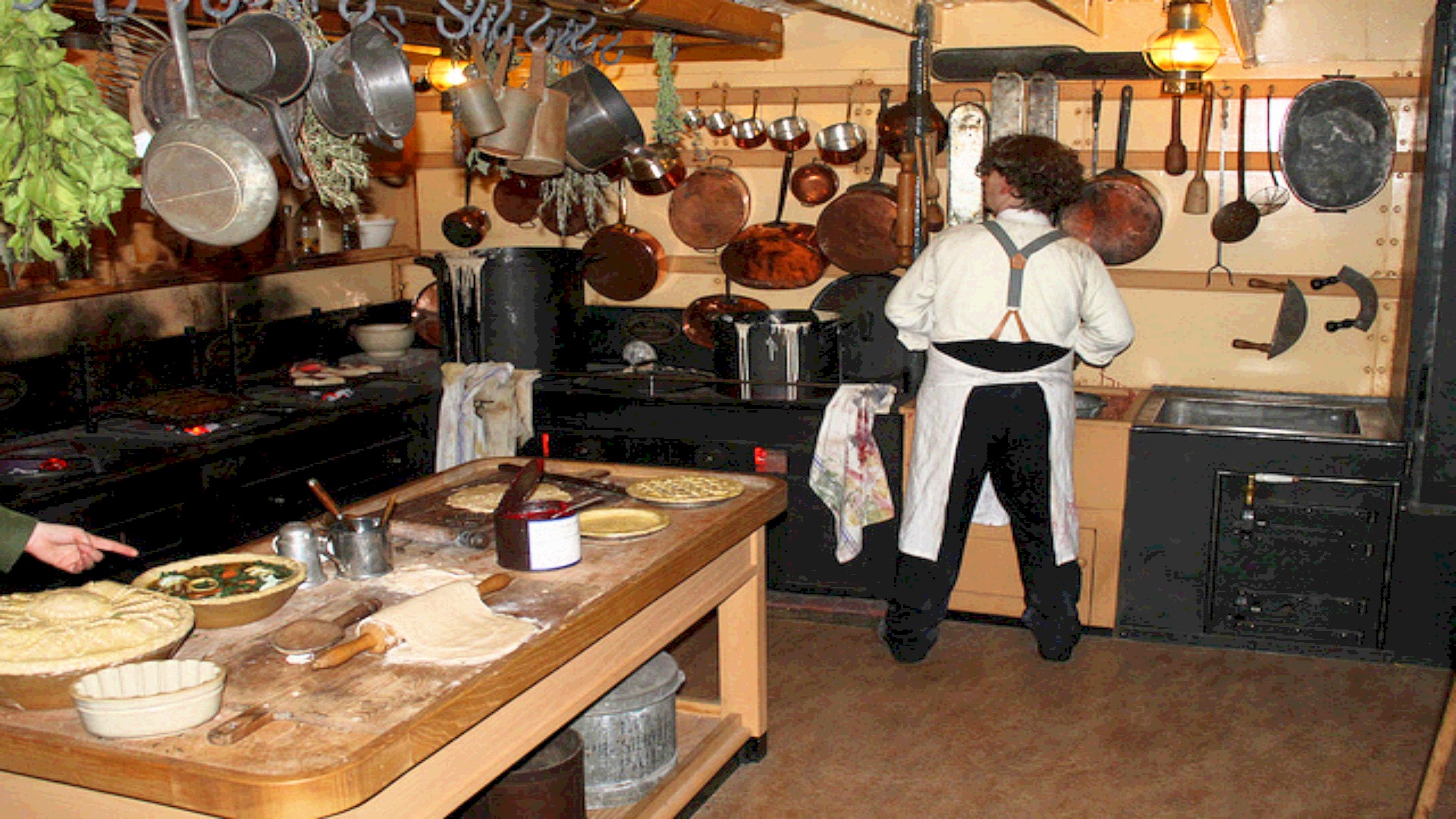}}
    \hfill
    \caption{Figure on the left (Fig. \ref{fig:histeq-example:a}) without HistEQ is applied, Figure on the right (Fig. \ref{fig:histeq-example:b}) is with HistEQ applied}
    \label{fig:histeq-example}
\end{figure}

There are three main groups of IPs that must be implemented.
First is the Deep Learning Unit (DPU).
The DPU and accompanying drivers are provided by Xilinx for accelerating the inference of Neural Networks (NNs).
Nearly all IPs provided by Xilinx, are designed to communicate using the AMBA 4 specification \cite{amba4}, primarily through the use of AXI-4 \cite{AXI-4} and AXI-Stream \cite{AXI-Stream} connections.
The DPU has three AXI connections, show in middle right of Figure \ref{fig:sys-overview}.
Two of which are 128 bits wide and are for receiving image data and model weights, and one connection that is 32 bits wide for fetching instructions.
The second IP group is HDMI.
The HDMI pipeline is composed of multiple IPs (depicted in bottom left of figure \ref{fig:sys-overview}), a physical interface IP (PHY), an HDMI receiver (RX) IP and an HDMI transmitter (TX) IP. 
These IPs together allow the ZCU104 to receive video from one physical HDMI connection (show in bottom of Figure \ref{fig:sys-overview}), and transmit video on another connection simultaneously.
The third IP group is the Video Enhancement Pipeline (VEP).
This pipeline allows for implementing image enhancements as a hardware pipeline.
The VEP and High-Definition Multimedia Interface (HDMI) use AXI-Streams in order to pass video between IPs.
AXI-Streams pass data one pixel at a time.
The VEP is able to achieve a very high throughput using minimal hardware through the use of deep pipelineing.
Due to LEAP's complexity, it is necessary to share video frames between the PL and PS.
The Operating System (OS), depicted in the top of Figure \ref{fig:sys-overview}, can only interact with data that has been memory mapped.
AXI-Streams are not memory mapped and must be converted to an AXI-4 connection in order to comunicate with the PS.
For this, the Video Direct Memory Address (VDMA) IP is used for both HDMI and VEP to map video to and from PS memory.
For the VEP, we developed a well-known image enhancement IP that implements the histogram equalization (HistEQ) to brighten dark images.
An example of this enhancement is shown in Figure \ref{fig:histeq-example}
The HistEQ IP is designed to work on different video resolutions. 
It has two raw inputs that each have a 12 bit input, allowing for enhancing video frames up to 1080p (1920 $\times$ 1080) resolution.
The video resolution can be configured by the PS by the use of the AXI-GPIO IP.
Typically, the AXI-GPIO IP is used to connect the physical inputs/outputs of the ZCU104, however it can also internally.
Twenty-five of the thirty-two AXI-GPIO output bits are utilized to control the HistEQ IP.
Twelve bits to control the number of rows, twelve bits for the number of columns and 1 bit to reset the HistEQ IP. 
The HistEQ IP must first be put in a reset state, the value for the rows and columns can be changed.
\textbf{\textit{While we are only using one image enhancement IP currently, we designed the VEP to be easily scalable via daisy-chaining enhancement IPs.}}
Just one AXI-GPIO is needed to configure multiple image enhancement IP(s).
The ZCU104 offers several high performance AXI-Slave connections to the PS memory.
The AXI-Slave connections are what allow for IPs in the PL to access memory in the PS address space.
The as already stated the DPU has three buses.
These buses run at 300 MHz, but some computation in the DPU is clocked at 600 MHz, which essentially allows the DPU to achieve double data rate.
The HDMI also runs at 300 MHz, but only uses one 128 bit bus for both writing and reading video frames. 
For 1080p video, there is around 2.1 million pixels. 
Each pixel is 24 bits.
At 60 frame per second, each video stream requires about 3 Gbits/s. 
The HDMI bus is able to transfer a theoretical maximum of 76 Gbits/s, which should be ample to handle two video streams.
The VEP is configured the same as the HDMI, except for the HistEQ IP.
The HistEQ IP is clocked at 150 MHz.
As mentioned before, the VDMA translates an AXI-Stream to and from PS memory.
To accomplish this, the AXI-Stream has to have a certain number of memory addresses reserved for writing and reading frames.
The amount of memory required can be adjusted, to have enough space for one or several video frames.
This can be taken advantage of to move a video frame received from HDMI to the VEP without the use of the processors in the PS.
How this is achieved, is by having the HDMI VDMA write a received video frame to a particular memory location in the PS and have the VEP VDMA read from the same memory location.
Where the VDMAs read/write video frames is controlled by the PS at runtime, however, there is one hardware requirement: both VDMAs must have a buffer of the same size.
For LEAP, we chose to set the frame buffer size of four frames for both VDMAs.
To generate the bitstream we use a makefile from the DPU-PYNQ \cite{dpu-pynq} project.
More information about DPU-PYNQ and why we are using it is explained in Section \ref{runtime-software}.
This makefile makes calls to Vitis \cite{vitis20221} to generate three files needed at compile time: The bitstream file (.bit), the hardware description file (.hwh), and a DPU linking file (.xclbin).
The bitstream file contains compiled versions of the IPs shown in Figure \ref{fig:sys-overview} that will be loaded into the PL a run time. 
The hardware description file gives the software additional information on what hardware that is required by the bitstream.
The DPU linking file is required by the DPU drivers for controlling the DPU IP.

\subsection{Compiling PYNQ Image} \label{compiling}

To accommodate LEAP's system integration, the PYNQ default image needs to be recompiled. 
The issue is the amount of Continuous Memory Allocation (CMA).
CMA is a very low level part of the Linux kernel.
CMA guarantees that there is a certain amount of continuous memory available.
CMA is what Direct Memory Access (DMA) in the PS uses to communicate with hardware in the PL.
CMA must be reserved at boot time in order to guarantee it will be continuous.
The publicly available PYNQ SD card image has a CMA of 512 MB.
If we were only using the DPU this would be sufficient \cite{dpu-cma-requirement}, but as we are using several IPs that require a large amount of CMA, the CMA must be enlarged. 
PYNQ provides documentation on how to compile its image \cite{pynq-sd-build}. 
At the time of writing, the latest version of PYNQ is v3.0, so that is what we are using.
PYNQ requires some tools and configuration in order to build an SD card image.
\textbf{\textit{To aid researchers compiling PYNQ, we created a Docker container, as part of our contribution (we'll provide our code as a weblink, and include it as a citation in this paper)}}.
We followed all the steps in the PYNQ documentation with one exception.
After cloning the PYNQ project, but before compiling, one file must be modified.
The file is located in PYNQ/boards/ZCU104/petalinux\_bsp/meta-user/recipes-bsp/device-tree/files/system-user.dtsi.
In the “reserved-memory” section, the “size” parameter must be changed from “0x20000000” to “0x40000000”.
This will double the CMA reserved at boot from 512 MB to 1024 MB.
Once this file is modified, the PYNQ image can be compiled.
Once the PYNQ SD card is created, it is burned to an SD card and booted off of.
As another contribution, we are providing a prebuilt PYNQ image.
It can be found in \cite{custom-pynq-img}.

\subsection{Runtime Software} \label{runtime-software}

PYNQ comes with drivers for interfacing with DPU, HDMI and VDMA, however it does come with python library for the DPU IP.
Instead of this, the DPU library is a separate project named: DPU-PYNQ \cite{dpu-pynq}.
After the ZCU104 is booted, the DPU-PYNQ library can be installed.
Using this library, it is relatively straight forward to submit jobs (models) to the DPU for inference.
Once the ZCU104 is booted, we configure the DPU, HDMI, and VEP IP groups.
For HDMI, we get the resolution and frame rate from the HDMI input.
We then use this information to set the HDMI output to the same resolution and frame rate.
For the VEP pipeline, we set the resolution to be the same as the HDMI resolution.
This way, we do not need to resize the image between the HDMI input and VEP pipeline.
To configure the DPU, we must first copy the .xclbin file to \/usr\/lib\/dpu.xclbin. 
If this is not done, it will cause the ZCU104 to crash, and will require rebooting.
Once the .xclbin file is copied, the .xmodel file can be loaded.
We created three ways methods of gathering, enhancing, and inferencing on a video signal.
Our reasoning for doing this is, different applications of LEAP may benefit from one method more than another (e.g. overlaying CNN results on the video signal).

\subsubsection{Asynchronous Inference}
\begin{figure}[h]
    \centering
    \includegraphics[width=0.48\textwidth]{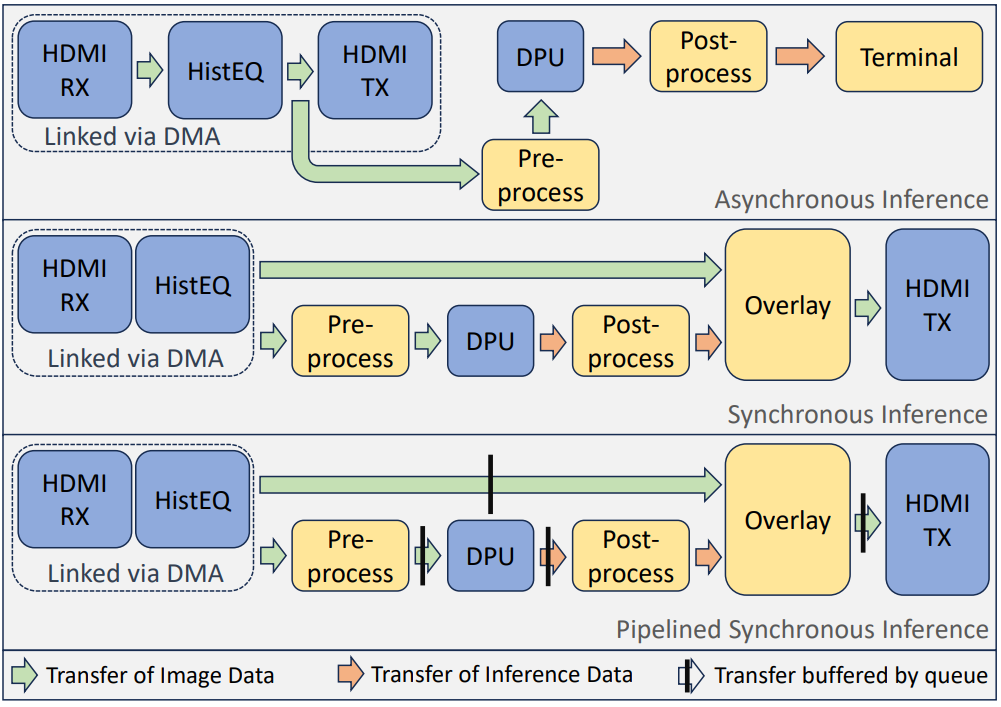}
    \caption{Our contribution of three different methods of performing inference: asynchronous inference, synchronous inference and pipelined synchronous inference.}
    \label{fig:inference-method}
\end{figure}

We can configure the HDMI receiver to store a video frame in the same location VEP is reading from.
VEP then writes the enhanced video frame to another location in memory that is read by the HDMI transmitter.
This allows the video frame to be transferred using hardware alone. 
For inferencing, the output of the VEP is read, preprocessed and sent to the DPU.
As soon as the DPU is finished inferencing, the output is logged to a terminal.
This is depicted in the top block of Figure \ref{fig:inference-method}.
The result of this implementation is the video processing and DPU inference run asynchronously.

\subsubsection{Synchronous Inference}
In some applications, it is desirable to display the prediction of a model on the video output (e.g. adding bounding boxes).
To overlay an image onto a video frame, we created a function in Python using openCV to modify a video frame.
As openCV is written in C, it is efficient and quick.
Unlike the DMA transfer, we are modifying the image being sent out of the ZCU104.
This limits us to only using DMA transfer between the HDMI receiver and the VEP.
First, a video frame is passed from the HDMI receiver to the VEP. 
The output of the VEP is read by software, preprocessed and passed to the DPU.
Once the DPU is finished processing and has made a prediction, the prediction is passed to our overlay function.
This function overlays the information onto the video frame (before it was preprocessed), and the modified video frame is sent to the HDMI transmitter.
This processing pipeline is synchronous to the DPU prediction.
The middle block in Figure \ref{fig:inference-method} illustrates how data is transferred between processing steps.
Unlike before, the frame rate of the HDMI output varies depending on how long it takes the DPU to inference.

\subsubsection{Pipelined Synchronous Inference}
The ZCU104 has four ARM Core A53s.
The synchronous inference as described above operates in a single process.
This limits the software to only running on one core.
By lifting this restriction, we expect to be able to achieve a higher throughput.
In order to fully utilize all the ZCU104's processors, a software based pipelining approach is used.
To do this, we split the workload into four parts, using four processes and three queues (with max depth of four items) to pass data between each process.
The placement of each queue is shown in the bottom block of Figure \ref{fig:inference-method}.
This way we could be processing four video frames at the same time instead of just one.
The first process reads a video frame and preprocess it, then send the result to the second process.
The second process passes the preprocessed video frame to the DPU and then sends the resulting video frame to the third process.
The third process would post process the result of the DPU and overlay the results on the video frame obtained from the VEP output.
Then it passes it to the fourth process.
The fourth process sends the frame to the HDMI transmitter.
The reason for grouping certain steps into one process is to load balance each process to take a similar time to other processes.
The groups are determined empirically by measuring the latency of each step.

\section{Experimental Setup} \label{benchmark}

\begin{figure}[H]
    \centering
    \includegraphics[width=0.45\textwidth]{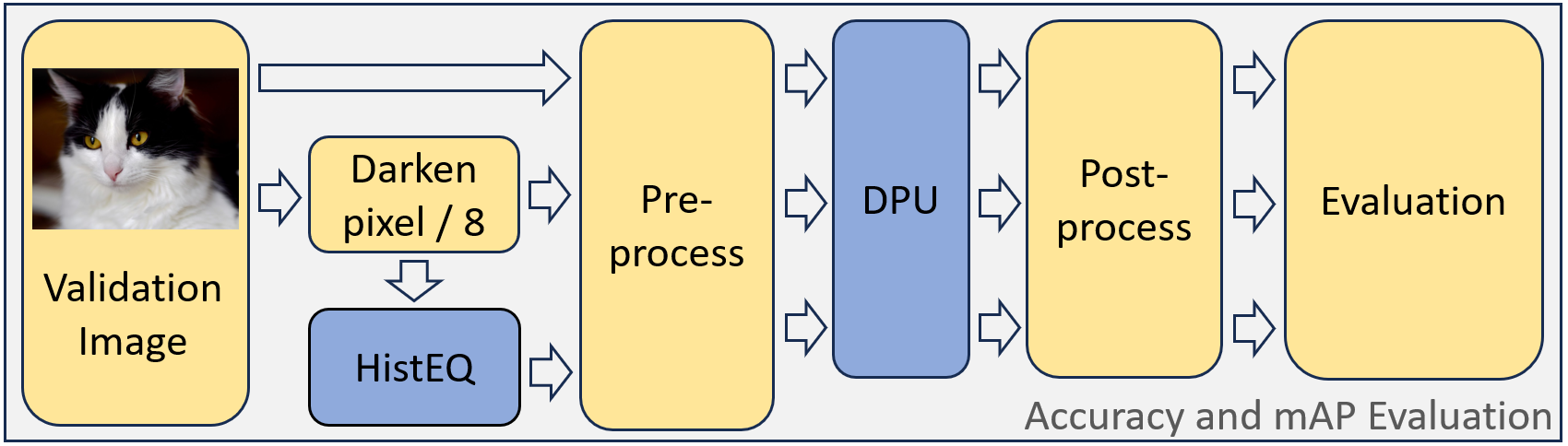}
    \caption{Setup for evaluating accuracy and mAP for Resnet50 and YOLOv3.}
    \label{fig:eval-method}
\end{figure}

To validate our methodology, we elected to use two types of CNNs, object detection and classification.
For objection detection we are using YOLOv3, from the Vitis-AI model zoo \cite{yolov3-model}, that has been trained on the Common Objects in Context (COCO) dataset \cite{cocodataset}.
For classification, we are using Resnet50, provided with the PYNQ-DPU project \cite{resnet50-model}, that was trained on the ImageNet dataset \cite{deng2009imagenet}.
We evaluated the performance of LEAP using three metrics, FPS for both models, mAP 50:95 for YOLOv3 and accuracy (top-1 and top-5) for Resnet50.
For FPS, we collected data using both the synchronous inference implementation and pipelined synchronous inference.
For the synchronous inference implementation, we measured the time latency of each step, such as the time required, to read a video frame from the VEP, inference time, post-processing time, etc.
We did not evaluate the FPS for the asynchronous inference implementation, as it is unaffected by models' size and complexity, and is only affected by the HDMI input frame rate which was held constants for all of our testing.
To objectively evaluate mAP and accuracy for Resnet50 and COCO, we used the validation datasets of Imagenet and COCO respectively.
We also want to show that our image enhancement is working, by artificially darkening images from the Imagenet and COCO validation datasets.
To darken the image, we divide each pixel by eight, making the image one eighth as bright as it was originally.
The evaluation methodology is illustrated in Figure \ref{fig:eval-method}. 
It should be noted that each test is carried out separately, then compared when all the tests have finished.
In Table \ref{tab:sync_time} it is shown that Resnet50 benefits marginally from using software pipelining, but YOLOv3 benefits by a significant amount.
The reason for this can be seen in Table \ref{tab:sync-time-breakdown}.
Resnet50 and YOLOv3 have different bottlnecks.
Resnet50's bottleneck is the write frame step, where YOLOv3's bottleneck is post-processing step.
By computing the steps across four processes, we run 4 steps in parallel.
\begin{table}[h]
    \caption{\small{Inference time for synchronous inference (SI) and pipelined synchronous inference (PSI).}}
    \label{tab:sync_time}
    \centering
    \begin{tabular}{|c|c|c|c|}
        \hline
        Model & SI Time & PSI Time & Difference \\
        \hline
        Resnet50 & 92ms & 83ms, & 9ms\\
        YOLOv3 & 334ms & 250ms & 84ms\\
        \hline
    \end{tabular}
\end{table}
\begin{table}[H]
    \caption{\small{Time of each step in synchronous inference.}}
    \label{tab:sync-time-breakdown}
    \centering
    \begin{tabular}{|c|c|c|c|c|c|c|}
        \hline
              & Read  & Pre-    &     & Post-   &         & Write \\
        Model & Frame & process & DPU & process & Overlay & Frame \\
        \hline
        Resnet50 & 1ms & 28ms & 13ms & $<$ 1ms & $<$ 1ms & 50ms \\
        YOLOv3 & 1ms & 43ms & 87ms & 153ms & $<$ 1ms & 50ms \\
        \hline
    \end{tabular}
\end{table}
\begin{table}[h]
    \caption{\small{Accuracy of Resnet50 and mAP of YOLO evaluated on Imagenet and COCO in dark environment with and without image enhancement applied. DS is short for Dataset.}}
    \label{tab:acc-map}
    \centering
    \begin{tabular}{|c|c|c|c|c|c|c|}
        \hline
              &        & Original    & Dark    & Dark DS  & \\
        Model & Metric & DS          & DS & + HistEQ & Increase \\
        \hline
        Resnet50 & Top-1 Acc. & 62\% & 26\% & 55\% & 29\% \\
        Resnet50 & Top-5 Acc. & 85\% & 49\% & 79\% & 30\% \\
        YOLOv3 & mAP 50:95 & 29 & 19.8 & 27.7 & 7.9 \\
        \hline
    \end{tabular}
\end{table}

For our accuracy and mAP metrics, Table \ref{tab:acc-map}, we can see once the image is darkend both Resnet50 and YOLOv3 performance are significantly reduced when compared to inferencing on the original datasets.
Once the HistEQ is applied, we see a significant jump in accuracy and mAP of Resnet50 and YOLOv3.

\section{Conclusion} \label{conculsion}
In this work, we provided a system integration to provide hardware/software integration of end-to-end methodology for intelligent CV applications on FPGA boards, leveraging Xilinx's PYNQ environment. 
We explored three methodologies for inferencing a CNN in real-time using Xilinx's ZCU-104 board.
To validate, our proposed methodology, LEAP, we used two CNN models with an image enhancement algorithm, histogram equalization, to enhance dark images.
We saw a frame rate of more than 10 frames per second and an accuracy increase as much as 30\% using our histogram equalization IP compared to not using it.
For promoting open research, we are providing complete code to recreate our work (we'll provide our code as a weblink, and include it as a citation in this paper).

\newpage

\bibliographystyle{ieeetr}
\bibliography{refs}


\end{document}